\documentclass[10pt, conference, letterpaper]{IEEEtran}

\usepackage{caption}
\usepackage{authblk}
\usepackage{enumitem}

\usepackage{comment}
\usepackage[utf8]{inputenc}
\usepackage{tensor}
\usepackage[cmex10]{amsmath} % Use the [cmex10] option to ensure complicance with IEEE Xplore (see bare_conf.tex)
\usepackage{physics}
\usepackage{amssymb,amsfonts}
\usepackage{mathtools,physics, bbm}

\usepackage{amsthm}
\newtheorem{theorem}{Theorem}

\newtheorem{remark}{Remark}
\newtheorem{proposition}{Proposition}
\newtheorem{definition}{Definition}
\newtheorem{myassumption}{Assumption}

\newtheorem*{Game*}{Game}
\newtheorem*{TokGen*}{TokenGeneration Phase}
\newtheorem*{TokVer*}{TokenVerification Phase}
\newtheorem*{Setup*}{Setup}
\newtheorem*{Query*}{Query}
\newtheorem*{Challenge*}{Challenge}
\newtheorem*{Guess*}{Guess}
\usepackage{breqn}
\newtheorem{subtheorem}{Theorem}[theorem]
\newtheorem*{note}{Note}
\usepackage{tikz}
\usetikzlibrary{quantikz}
\DeclareCaptionType{mycapequ}[][List of equations]
\usepackage{multirow}
\usepackage{graphicx}

\usepackage{tikz}
\usetikzlibrary{quantikz}

\usepackage{nameref}

\usepackage{graphicx}
\usepackage{url}
\usepackage{cite}
\usepackage{hyperref}
\usepackage[capitalise]{cleveref}
\hypersetup{
    hidelinks = true
}
\begin{document}
\title{Entanglement-assisted authenticated BB84 protocol}

\author[1]{Pol Julià Farré\textsuperscript{*}}
\author[2]{Vladlen Galetsky}
\author[2]{Soham Ghosh}
\author[2]{Janis Nötzel}
\author[1]{Christian Deppe}

\affil[1]{Institute of Communications Engineering, Technische Universität Braunschweig, Braunschweig, Germany}
\affil[2]{Chair of Theoretical Information Technology, Technical University of Munich, Munich, Germany}

\twocolumn[
\begin{@twocolumnfalse} % forces full-width mode explicitly

\maketitle

\vspace{-1em}
\begin{center}
\textsuperscript{*}Corresponding author: pol.julia-farre@tu-bs.de
\end{center}

\vspace{1em} % some spacing after

\end{@twocolumnfalse}
]

\begin{abstract}
    In this work, we present a novel authenticated Quantum Key Distribution (QKD) protocol employing maximally entangled qubit pairs. In the absence of noise,  we securely authenticate the well-known BB84 QKD scheme under two assumptions: first, adversaries cannot simultaneously access pre-shared and non-pre-shared secret classical information, and second, adversaries cannot simultaneously access pre-shared secret  classical information and  quantum memories held by legitimate parties. The main strength of this noiseless result is that access to all secretly pre-shared classical information is insufficient for breaching our scheme. Additionally, our protocol desirably allows for pre-shared secrecy reusage, leading to secret key growing.
    
    In order to address noise, we simulate a photonic implementation of our scheme, together with a storage model that aims to replicate the performance of cavity-enhanced Atomic-Frequency Comb (AFC) memories. Two methods are used to distinguish authentic entities from forgery attempts: on the one hand, a statistical approach is used after calibration of its defining parameter $\mu$.  Alternatively, a Deep Neural Network (DNN) is designed and trained to learn the underlying different structure of that input data coming from adversaries in comparison to that one coming from legitimate parties.  Both methods achieve a correct classification rate larger than 0.80 for memory storage time of 150 $\mu \mathrm{s}$ and a 1 $\mathrm{km}$ distance between  parties.

\end{abstract}

\begin{IEEEkeywords}
Authenticated QKD, BB84 protocol, AFC memory, entanglement-assisted authentication, two-factor authentication, secret pre-sharing, man-in-the-middle attack.
\end{IEEEkeywords}

\section{Introduction}

Quantum Key Distribution (QKD) gained significance following the derivation of Shor’s algorithm \cite{shor}, which introduced a major security threat: if sufficiently large and efficient quantum computers become feasible, Shor’s algorithm can break widely used public-key cryptosystems such as the RSA cryptosystem \cite{RSA}, the Diffie–Hellman key exchange protocol \cite{diffie_hellman}, and elliptic curve cryptography \cite{elliptic_curves}, which differ in design but are all vulnerable to quantum attacks. Within the new paradigm introduced after Shor’s derivations, QKD enables the secret distribution of a shared key between two parties. Unlike classical approaches to the same task—such as post-quantum cryptography and physical layer security—QKD offers unconditional security, grounded in the principles of quantum mechanics. Post-quantum cryptography has been introduced in a general overview \cite{pqc}, reviewed in  \cite{review_postquantum} and in \cite{review_post_Q}, and surveyed in a book covering its mathematical foundations and implementation aspects \cite{bernstein_book}. Relevant discussions on physical layer security include the overview in \cite{overview_PLS}, the review in \cite{physical_layer}, a study on optical fiber networks \cite{PSL_2.0}, and a comprehensive book on quantum communication networks \cite{QCN_BOOK}.
The reliability of the quantum-mechanical principles underpinning QKD has been substantiated by experimental demonstrations of the Heisenberg uncertainty principle with fullerene molecules \cite{exp_unc_prin_1} and in the context of entanglement witnessing \cite{exp_unc_prin_2}, as well as by loophole-free violations of Bell inequalities, demonstrated with electron spins \cite{exp_bell_1} and superconducting qubits \cite{exp_bell_2}.

Two major variants of QKD were introduced through the seminal proposals in \cite{bennet_brassard} and in \cite{e91}. The first gave rise to prepare-and-measure protocols, while the second led to entanglement-based protocols. Since then, many variations have been proposed to improve security and practicality. An overview of QKD and its development over recent decades is provided in \cite{review_QKD_2.0}, with a more recent review available in \cite{other_review_qkd}. A comprehensive survey is presented in \cite{survey_QKD_2.0}. Continuous-variable QKD has been reviewed in \cite{cont}, experimentally demonstrated in \cite{CV_QKD}, and further developed by leveraging two-way quantum communication in \cite{cv_qkd_2}. Within discrete-variable protocols, decoy-state QKD forms a significant subclass. The concept was first proposed in \cite{decoy_seminal} and expanded in \cite{decoy}. A relevant experimental realization using polarized photons is reported in \cite{decoy_realization}, and improvements addressing photon-number splitting attacks are detailed in \cite{decoy_bs_attack}. It is also important to emphasize that current QKD prototypes already operate in optical fibers at intercity ranges, covering distances from 45 \cite{qkd_tokyo} to 78 km \cite{jena}. This, combined with the fact that satellite-based QKD, as reported in \cite{satellite_qkd}, can operate over distances exceeding 4000 km, highlights that QKD has progressed beyond its initial stage, where it was regarded merely as a proof-of-concept technology \cite{proof_of_concept}.

In parallel to QKD, we introduce the concept of entity authentication. Entity authentication can be based on something an entity uniquely \textit{has} (token-based authentication \cite{token_2}), something an entity uniquely \textit{knows} (e.g., password-based authentication \cite{passwords_1}), or something an entity uniquely \textit{is} (biometrics-based authentication \cite{biometric_review}), and like other foundational security building blocks, it has traditionally been grounded in classical techniques. However, with the advent of quantum technologies, the field of entity authentication is expected to advance, potentially enhancing resilience against current and emerging threats.

A comprehensive review of quantum entity authentication methods is provided in \cite{review_quantum_authentication}, highlighting recent advancements such as Quantum Physical Unclonable Functions (QPUFs), extensively reviewed in \cite{review_QPUFs}, and which stem from the earlier Classical Physical Unclonable Functions (CPUFs) \cite{PUFsreview}. While QPUFs show promise, they are not yet practical for widespread use. In contrast, CPUFs have demonstrated practical viability \cite{cpuf_vehicle}, although they have been found vulnerable in certain cases \cite{xor}, and, unlike (some of) their quantum counterparts, require a trusted third party.

To conclude this introduction, we emphasize the interconnection between QKD and entity authentication. While sometimes overlooked in the literature, QKD fundamentally relies on the critical assumption that the involved parties are securely authenticated \cite{basics_qkd_auth_1}. This authentication, essential for subsequent secure communication, can be achieved through classical or quantum methods, with our work serving as an example of the latter. Importantly, given that authentication schemes often rely on potentially violated assumptions (e.g., trusted parties or secrecy of pre-shared information), multifactor schemes are commonly employed to reinforce security robustness \cite{multifactor_1}.
\subsection*{Our main contributions}

In this work, we define a new protocol that in the noiseless case provides two-factor information-theoretically secure  authentication for the  BB84 QKD scheme \cite{bennet_brassard}. That is, it is  mathematically ensured that breaching our protocol requires two independent actions: access to pre-shared secret classical information and either access to non-pre-shared secret classical information or to quantum memories held by legitimate parties. 

Our work presents a new contribution to QKD authentication using quantum methods, addressing certain vulnerabilities found in a recent proposal \cite{main_citation}. Similar to \cite{pseudorandom} and \cite{no_public_bases}, the cited authors propose authenticating the quantum channel within the BB84 protocol, rather than authenticating each subsequent use of the public classical channel. While their scheme is resistant to several security threats, it remains vulnerable to the leakage of pre-shared secret classical information, a weakness also common to the prevalent classical alternatives. In contrast, our approach leverages quantum entanglement and exploits non-pre-shared secrecy, ensuring that access to pre-shared secret classical information alone is insufficient to compromise security. Furthermore, the proposal in \cite{main_citation} lacks a discussion on the security implications of reusing the secretly pre-shared key or on a mechanism that guarantees key chaining across different sessions. In contrast, our scheme enables pre-shared key recycling, a crucial feature for having practical authenticated QKD.

In order to assess our scheme noise adaptations, we simulate qubit transmission via photonic optical fiber channels and qubit storage using cavity-enhanced AFC memories. By outlining the evolution of key performance metrics for the required hardware over the past years, we acknowledge the potential of our proposed designs and implicitly  draw  future perspectives based on this  progress.

\subsection*{Outline}
Section \ref{preliminaries} includes all relevant definitions, and states the notation  followed throughout the article. Section \ref{reviewBB84} provides a comprehensive review of the BB84 protocol, and  Section \ref{main} includes the noiseless definition of our scheme, and its noise adaptations. In Section \ref{sec analysis}, we analyze the security of the presented protocol, and in  Section \ref{benchmark}, we highlight the main strengths of it, as well as its potential drawbacks. Subsequently we proceed to contextualize  our work by comparing it with the principal related proposals found in the state of the art. In Section \ref{realistic}, we describe in detail the noise model that we adopt for our simulations, and in Section \ref{results}, we present and discuss the corresponding outcomes. Section \ref{conclusions} constitutes a concise summary of the text, including our main findings, and Section \ref{further_research} serves us to outline avenues for future research as well as to conclude the article.

\begin{note}
  This article constitutes a revised and extended version of a seminal work presented at the European Wireless Conference 2024 \cite{EW}.
\end{note}

\section{Preliminaries}
\label{preliminaries}

In this section, we first provide a brief glossary of key terms related to security, adapting existent definitions in the literature. Following that, we establish the notation that is used throughout the article.

\subsection{Glossary}
\label{glossary}

Two definitions originate from formal logic: a statement is valid if it is provable (the completeness property), and only a provable statement is valid (the soundness property). These definitions have been adapted and adopted  in various works, such as \cite{qmoney_security_def} and \cite{main_citation}, to establish security for their authentication systems. We analogously define:

\begin{definition}
\label{completeness_def}
    $\epsilon_c$-completeness:
    
    Property only owned by those authenticated QKD protocols that, in the absence of malicious parties and unfavorable environmental conditions, are aborted with a probability smaller than or equal to $\epsilon_c$.
\end{definition}

\begin{definition}
  \label{soundness_def}
   $\epsilon_s$-soundness:

  Property only owned by those authenticated QKD protocols in which, under no protocol abortion, none of the legitimate parties involved can be misled, with probability greater than $\epsilon_s$, about the identity of their key-sharing counterpart.
\end{definition}

\begin{remark}
\label{soundness clarification}

On the different meanings of soundness:

It is important to clarify that, within a framework where the authentication layer for QKD is already assumed, the soundness property  is typically used  with a distinct meaning \cite{security_definition_1, security_definition_2}. That is, in such case, soundness refers to that property only owned by those QKD protocols that correctly and secretly distribute a key between the legitimate parties involved. The scenario that concerns us is somewhat the converse, i.e., we assume correctness and secrecy of the distributed key, and we set, with our definition, an entity authenticity requirement.

\end{remark}

Together, the two previous definitions aid the following characterization of what a secure authenticated QKD protocol must fulfill.

\begin{definition}
 \label{security_def}
    $(\epsilon_c,\epsilon_s)$-security: 
 
   Property only owned by those authenticated QKD protocols that are $\epsilon_c$-complete and $\epsilon_s$-sound.
   
\end{definition}

Lastly, we define two well-established types of  security attacks.

\begin{definition}
    Impersonation attack:

   Type of attack attempting to breach security within a certain protocol by forging the identity of a legitimate entity. 
\end{definition}

\begin{definition}
    Man-in-the-middle attack:

    Type of attack in which an adversary accesses transferred inputs between  different nodes of a communication protocol, attempting to break its security.
\end{definition}

\subsection{Notation}

During the whole paper, we use the Dirac notation to represent quantum states. In addition, we denote quantum gates  as capital letters.

The following definitions serve us as examples of such convention and allow us to present the two non-trivial single-qubit gates that appear in this work:

We introduce the well-known unitary operators $X$ (Pauli-$X$) and $H$ (Hadamard gate) by describing their action  onto the computational basis of $\mathcal{H}=\mathbb{C}^2$ as 

\begin{equation}
  X \ket{j} = \ket{j \oplus 1} \hspace{0.2cm} \& \hspace{0.2cm} H \ket{j} = \frac{1}{\sqrt{2}} \Big( \ket{0}+ (-1)^{j}  \ket{1} \Big),
\end{equation}
where $j \in \{0, 1\}$.

As a final remark, all quantum circuits shown are in agreement with the common formalism used in quantum computation. Hence, their time axis runs from left to right, as well as quantum gates are enclosed within boxes. Measurements are taken at the very end of the circuit and, unless otherwise stated, occur on the computational basis.

\section{The BB84 protocol at a glance}
\label{reviewBB84}

In this section, we revisit the BB84 protocol\cite{bennet_brassard} for the sake of  self-containment.

\subsubsection{Secret key generation and encoding}

The BB84 protocol defines two communication parties: Alice, the sender, and Bob, the receiver. Alice randomly\cite{TRN} generates a secret binary key in order to end up sharing a portion of it with Bob\footnote{Such shared randomness can then be exploited to exchange a message with perfect secrecy\cite{one-time-pad_shannon}.}.  Each bit of the key generated by Alice is encoded in the state of a qubit. Bits with value "0" are encoded in either the state $\ket{0}$ or the state $\ket{+} \equiv \frac{1}{\sqrt{2}}(\ket{0} + \ket{1})$, with equal probability. Conversely, bits with  value "1" are encoded in either the state $\ket{1}$ or the state $\ket{-} \equiv \frac{1}{\sqrt{2}}(\ket{0} - \ket{1})$, also with equal probability. This encoding strategy is commonly referred to as encoding a bit in the $Z$ or $X$ basis, respectively.

\subsubsection{Key transmission and decoding}
\label{bases}
After the initial secret key has been encoded in a qubit string, Alice sends it to Bob, who must measure each qubit either in the $Z$ or $X$ basis out of a uniformly  distributed random choice that he records. Following this action, both Alice and Bob publicly acknowledge their choice of bases and, in those cases where the basis coincides, noiseless quantum theory ensures coincidence between the corresponding  bits obtained out of measurement.  This allows for a creation of a shared secret key of an expected size being half of the length of the originally encoded bitstring\footnote{Such key sifting could be avoided if Bob stored the received qubits until Alice communicates her  preparation bases choice \cite{BB84_with_memories}. In this manner, Bob could replicate Alice's choices in his measurements and fully recover the initial bitstring generated by her. However, to the best of the authors' knowledge, this is currently not feasible due to the fragility of quantum memories.}.

\subsubsection{Information reconciliation and privacy amplification}

Although the noiseless quantum theory validates the described protocol and allows for a rigorous proof of its security, real-world physical scenarios inevitably introduce errors from various sources, resulting in noisy systems. The authors of \cite{bennet_brassard} recognized this and later proposed the information reconciliation \cite{info_reconciliation} and privacy amplification \cite{priv_amplification} algorithms, which when applied sequentially, allow for error correction without secrecy leakage.

\begin{remark}
On the security proofs of the BB84 protocol:

Relevant security proofs that incorporate the two mentioned algorithms can be found  both in \cite{proof_bb84_1} and \cite{proof_bb84_2}, where the Quantum Bit Error Rate ($QBER$) needs to be upperbounded\footnote{The first cited authors specify that a maximum $QBER$ of approximately 11$\%$ is required, while the second cited work improved this result, raising the threshold to 12.4$\%$.}. However, as detailed within the review on quantum cryptography in \cite{review_QKD_2.0} and exemplified by the attack on commercialized devices derived in \cite{q_hacking_1}, there is a gap between the theoretic security of the BB84 protocol, as well as QKD protocols in general, and that of their available implementations. This fact, mainly caused by assumptions made on the involved hardware devices such as photon sources and detectors, hinders still today an official consensus and certification for commercialized QKD implementations. Recent progresses on device-independent QKD constitute a promising line of research in this field, as reviewed in \cite{device_independent}.
\end{remark}

\subsubsection{Eavesdropping resilience}
\label{check_eave}

A desirable property of the protocol we are describing is its ability to easily detect an intercept-resend attack \cite{intercept-resend}. This can be achieved by having Alice and Bob randomly split the initially generated key into two sets immediately after acknowledging the choice of bases. One set continues with the QKD protocol, while the other, comprising a small portion of the total key, is used to check for high correlation rates. Such high rates can only be achieved if no \emph{eavesdropper} intercepted and acquired information from the transmitted qubits \cite{bb84_eave}.

\section{Entanglement-assisted authenticated BB84 protocol}
\label{main}

In this section, we begin by stating our noiseless derivation of a scheme that authenticates the BB84  protocol.  Subsequently, we display its noise adaptation.

\subsection{Noiseless protocol definition}
\label{noiseless protocol}

\subsubsection{Setup phase}

 \label{proposal} 
\begin{itemize}
    \item \textbf{1st step:} Our scheme starts  by requiring Alice and Bob to spatially coincide in time before  \emph{a series of} $n$ QKD \emph{sessions} commence.

    \item \textbf{2nd step:} Alice and Bob are provided with $2n\lambda$ qubits that undergo, in pairs, the first stage of the circuit described in Fig.\,\ref{circuit}. A bitstring $F_{\mathrm{Choices}}$, only owned by Bob, encodes  $\lambda$ uniform random choices, $\mathbf{I}$ ("0") or $X$ ("1"), characterizing all the circuits employed:  $F_{\mathrm{Choices}}$ is concatenated $n$ times. 

\begin{figure}[h!]
\centering
\includegraphics[width=0.4\textwidth,clip]{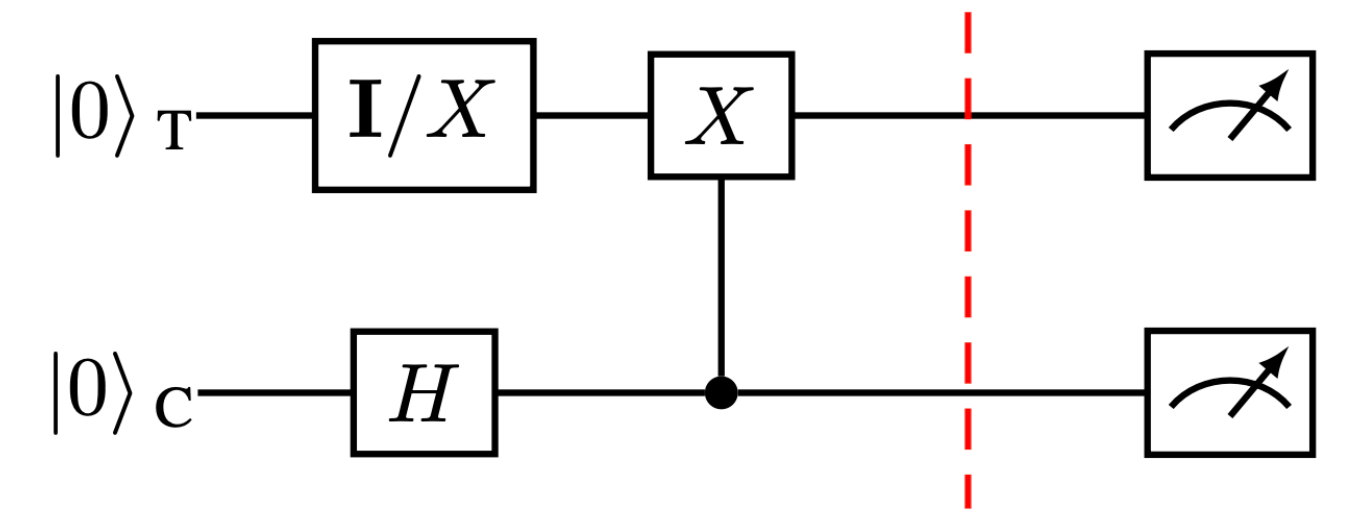}

\caption{On the left side of the red dashed line, entanglement generation occurs. When applying $\mathbf{I}$ to the target register (T), the maximally entangled pair $\frac{1}{\sqrt{2}}\big( \ket{0}_{\mathrm{C}} \otimes \ket{0}_{\mathrm{T}} + \ket{1}_{\mathrm{C}} \otimes \ket{1}_{\mathrm{T}}\big)$ is generated, while the maximally entangled state $\frac{1}{\sqrt{2}}\big( \ket{0}_{\mathrm{C}} \otimes \ket{1}_{\mathrm{T}} + \ket{1}_{\mathrm{C}} \otimes \ket{0}_{\mathrm{T}}\big)$ is created otherwise. After the two registers are  separated (red dashed line), Alice measures the register C and Bob measures the register T.}
\label{circuit}    
\end{figure}

\item  \textbf{3rd step:} Alice and Bob secretly pre-share a string \( K_{\mathrm{Positions}} \) consisting of \( \lambda \) integers, which encode positions chosen uniformly at random within a \( 2\lambda \)-qubit string\footnote{Notice that such integer-valued string can be encoded within a bitstring containing $\lambda$ "0"'s and $\lambda$ "1"'s. We propose to sample it by repeatedly sampling random bits until an even and balanced string of desired length is obtained.}.

 \item \textbf{4th step:} Alice stores the complete set of \( n\lambda \) maximally entangled pairs until both parties are ready to generate a shared key. Bob moves to the location where the quantum channel between the two parties terminates.

\end{itemize}

\subsubsection{Execution phase}

\begin{itemize}

\item \textbf{5th step:} At  each round, Alice sends $\lambda$ target qubits, what we label as authenticating (AU) qubits, embedded within a $\lambda$-qubit string used for QKD, and at the agreed positions $K_{\mathrm{Positions}}$.

\item \textbf{6th step:} At each round, Bob measures those $\lambda$ qubits allocated to QKD on randomly chosen bases, as stated within  the BB84 protocol. With regard to the remaining  $\lambda$ qubits, the AU ones, Bob measures them on the computational basis, in an order consistent with that of $F_{\mathrm{Choices}}$. He then aggregates the obtained outcomes into a bitstring $F_{\mathrm{Bob}}$. Alice  also measures her $\lambda$  control qubits (those entangled with the ones Bob measured, and in an analogous order), and  aggregates the results into $F_{\mathrm{Alice}}$.

\item \textbf{7th step:} Let us denote the bitstring encoding Alice's (Bob's) preparation (measurement)  bases choice within the BB84 scheme as $B_{\mathrm{Alice}}$ ($B_{\mathrm{Bob}}$). In this step Alice publicly communicates the bitstring 

\begin{equation}
    \Tilde{B}_{\mathrm{Alice}} \equiv B_{\mathrm{Alice}}\oplus F_{\mathrm{Alice}}
    \label{ecrypt1}
\end{equation}
 
 to Bob, who will assume 
 
 \begin{equation}
     \Hat{B}_{\mathrm{Alice}} \equiv \Tilde{B}_{\mathrm{Alice}} \oplus F_{\mathrm{Bob}} \oplus F_{\mathrm{Choices}}
     \label{assume1}
 \end{equation}
  as the actual preparation bases chosen by Alice when proceeding with the BB84 scheme. Analogously, Bob publicly communicates the bitstring

 \begin{equation}
    \Tilde{B}_{\mathrm{Bob}} \equiv B_{\mathrm{Bob}}\oplus F_{\mathrm{Bob}} \oplus F_{\mathrm{Choices}}
    \label{ecrypt2}
 \end{equation}
 
  to Alice, who will assume 

  \begin{equation}
      \Hat{B}_{\mathrm{Bob}} \equiv \Tilde{B}_{\mathrm{Bob}} \oplus F_{\mathrm{Alice}} 
      \label{assume2}
  \end{equation}
  
   as the actual measurement bases chosen by Bob when proceeding with the BB84 scheme. 

\item \textbf{8th step:} In the last step (see Fig.\,\ref{Full_sch} for a full protocol illustration), similar to the final steps outlined in the original BB84 protocol, Alice and Bob compare their bases choice, obtain a sifted key and check whether the Quantum Bit Error Rate (QBER) for a randomly selected fraction $\xi$ of such  key is $0$. If the QBER is $0$, Alice and Bob continue with the BB84 scheme and, otherwise, they abort the protocol.

\begin{figure*}
\centering
\includegraphics[width=1.0\textwidth,clip]{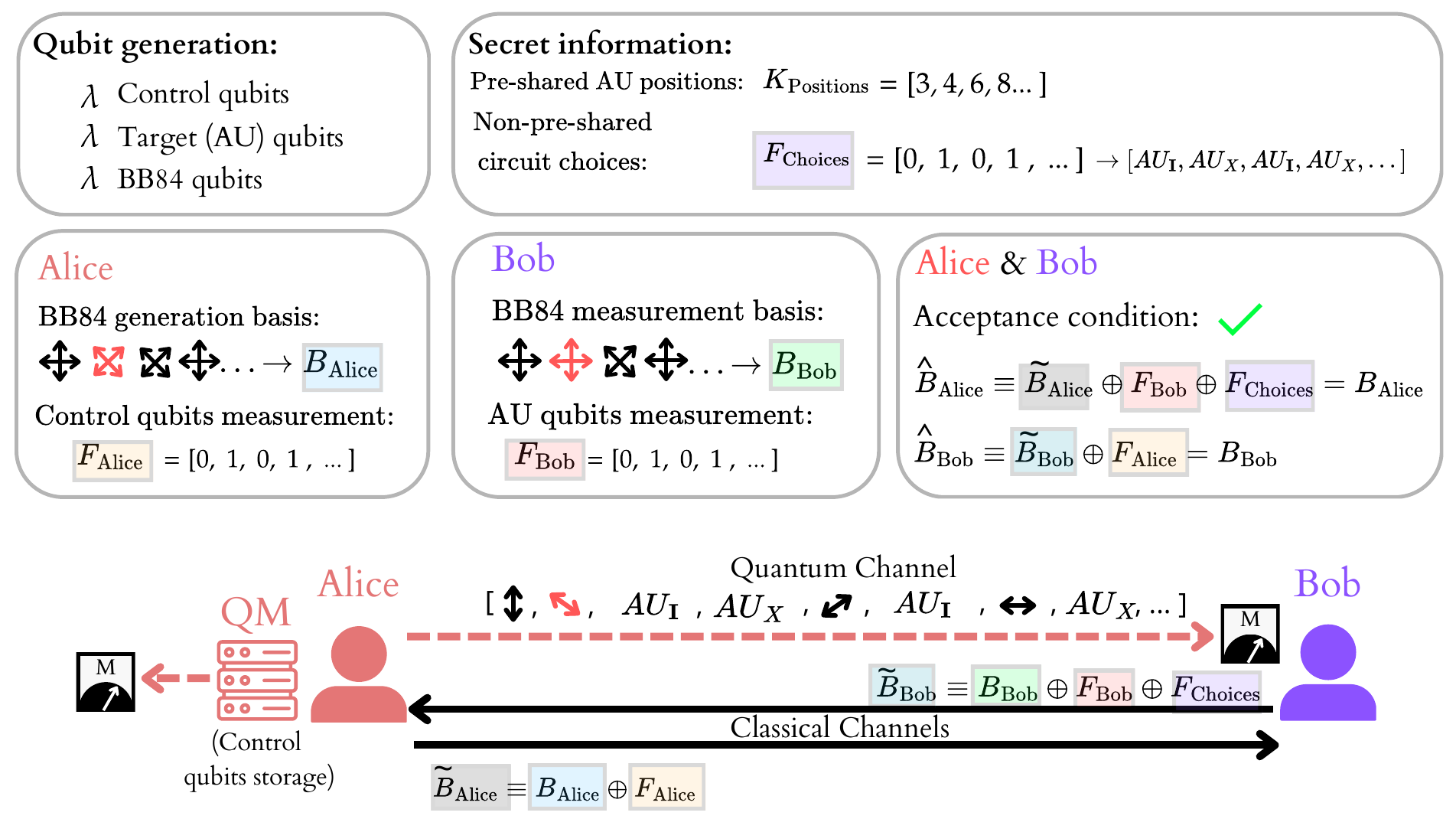}

\caption{Entanglement-assisted authenticated BB84 protocol.  We denote the quantum memories available at Alice's side as QM. For the BB84 qubits, a basis in red indicates a mismatch between the chosen basis for preparation and the basis used for measurement.}
\label{Full_sch}    
\end{figure*} 

\end{itemize}

\subsection{Noise adaptation}
\label{Noise_adaptation}

Due to the inevitability of errors arising from quantum gates, transmission, storage and readout, our  definitions for accepting each legitimate party lacks robustness. Therefore, we propose two alternative approaches with the objective of devising a more suitable protocol for a practical implementation.

\subsubsection*{Acceptance condition a)}

According to this definition, the two parties involved are mutually verified  if and only if the QBER observed within the eavesdropping-check stage of the BB84 protocol fulfills 

\begin{equation}
    1 - QBER \geq \mu,
\label{eq:r01}
\end{equation}
for some $\mu > \frac{1}{2} + \frac{1}{\sqrt{2\xi \lambda}}$.

\subsubsection*{Acceptance condition b)}

At the eavesdropping check, a binary string \( S \) has to be created, where each position takes the value "0" if the two compared bits coincide and a "1" otherwise. This string is then processed by a trained Deep Neural Network (DNN) \cite{sze2017efficient} specialized in solving the binary classification problem. That is, the DNN determines whether the input bitstring $S$ corresponds to two legitimate parties or if, instead, the protocol is aborted. 

\begin{remark} 
On a motivation for our second  noise adaptation:

One motivation for using a machine learning algorithm for this task is that the string \( S \) generated by  legitimate parties may exhibit a much more complex structure compared to those generated by  malicious parties, specifically: those randomized ones. The first proposed acceptance condition only captures that the mean of the string $S$ is typically lower than \(\frac{1}{2}\) for authentic entities. Nonetheless, specific setup choices may induce additional features to such string. For instance, the string of qubits measured at verification may be  non-uniformly  affected by noise (e.g., because of the time arrow). 
\end{remark}
 
\section{Security analysis}
\label{sec analysis}

We divide our security analysis into two parts. Within the first one, we discuss the security properties of our scheme when assuming ideal noiseless conditions, and in the second part, we analogously proceed with our two different adaptations to quantum noise conditions.

The limitations on the power of the adversary party are
crucial and must be clearly acknowledged in order for the
displayed security proofs to be meaningful. In the noiseless case, our adversarial model is characterized by the two following assumptions:

\begin{myassumption}
    Adversaries are not able to simultaneously access pre-shared and non-shared secret classical information. 
\end{myassumption}

 \begin{myassumption}
Adversaries are not able to simultaneously access pre-shared secret classical information and quantum memories held by legitimate parties. 
 \end{myassumption}

When taking in consideration noisy conditions, instead, the adversary is randomized.

\subsection{Security of our noiseless protocol definition}

Under the assumed adversarial model and noiseless conditions, our  protocol is $\Big(0, \big(\frac{3}{4}\big)^{\xi \lambda}\Big)$-secure  (see Proposition \ref{proposition_completeness} and Theorem \ref{theorem_security} in Appendix \ref{appendix_proofs}). The two assumptions made, become apparent in our security proofs (see Appendix \ref{appendix_proofs}), where they are explicitly exploited. Accordingly, we show resistance against attacks that exhaust all the adversarial resources allowed within our model, including theft of stored quantum systems and the strongest form among man-in-the-middle schemes, i.e.,  the straightforward man-in-the-middle attack \cite{straightforward}.

While our scheme possesses desirable theoretical properties, we acknowledge that some of the considered adversarial strategies, e.g., intercept-resend attacks \cite{intercept-resend}, can easily lead to protocol abortion. That is, such attacks can serve as means for jamming \cite{jamming_survey}. Additionally, it is important to emphasize that our security analysis is inherently tied to the specific adversarial model we assume. Although this assumption is often implicit in many security analyses, given its self-evident nature, we believe that it is important to recognize that our scheme remains vulnerable to unforeseen threats.

\subsection{Security of our protocol noise adaptations}

\subsubsection{On the security of the acceptance condition a)}

Against a randomized attacker, our first noise-adapted protocol is $\gamma$-sound, where we define:

\begin{equation}
    \gamma \equiv \frac{1}{2\xi \lambda}\frac{1}{\Big(\mu-\frac{1}{2}\Big)^2}.
\end{equation}
That is, a randomized attacker has a probability of success $p_{\mathrm{A}}$ fulfilling (see Proposition \ref{proposition_noise}) 

\begin{equation}
    p_{\mathrm{A}} \leq \gamma, 
    \label{mu_equation}
\end{equation}
which establishes a non-trivial bound due to the specified constraint when choosing  \(\mu\) (see Section \ref{Noise_adaptation}).

With regard to completeness, whether the protocol will be aborted or not in the absence of malicious parties is now dependent on the noisy hardware implementation.

\subsubsection{On the security of the acceptance condition b)}

Regarding our second noise adaptation, its security is heavily dependent on the training process of the employed DNN, which functions as a black box. Thus, in this case the security is of  heuristic nature and does not fit our $(\epsilon_c, \epsilon_s)$-security framework.

\section{Assessment and contextualization}
\label{benchmark}

In this section, we highlight the key strengths and limitations of the proposed authentication scheme, which helps mitigate impersonation attacks within the QKD framework. We then benchmark our proposal, emphasizing the advantages of using quantum technologies over classical resources for QKD authentication, and compare the features of the presented protocol with the current state of the art in quantum-authenticated QKD. To complement this, Table \ref{explanatory_table} explicitly compares key features of our protocol with the main alternatives relying on pre-shared secrecy.

\begin{table*}
  \centering
  \caption{Comparison among the security features of our proposed entanglement-assisted authenticated BB84 protocol, two other relevant schemes in the state of the art, and
  MAC-based solutions, already initially suggested in \cite{bennet_brassard}.}
  \renewcommand{\arraystretch}{1.2}
  \begin{tabular}{|c|c|c|c|}
    \hline
    \textbf{Attack goal} & \textbf{Eavesdropping} & \multicolumn{2}{c|}{\textbf{Impersonation}} \\
    \cline{3-4}
    \hline \textbf{Attack strategy} & \textbf{Intercept-resend} &  \textbf{Man in the middle} & \textbf{Access to pre-shared secrets}\\
    \hline
    MAC-based solutions & Secure \cite{intercept-resend}  & Secure & Not secure \\ \hline
     H. Park et al. \cite{main_citation} & Inherited security   &  Secure  & Not secure 
     \\ \hline
     Q. Jia et al. \cite{key_update} & Inherited security   &  Secure  & Not secure
     \\ \hline
     Our proposal & Inherited security   & Secure (Theorems\, \ref{theorem_theft} and \ref{theorem_classicical_theft}) & Secure (Theorem \ref{theorem_theft})\\ \hline
  \end{tabular}
  \label{explanatory_table}
\end{table*}

\subsection{Desirable properties}
\label{desirable_properties}

Under noiseless conditions, our proposal offers a two-factor information-theoretically secure authenticated QKD scheme (see Proposition \ref{proposition_completeness} and Theorems \ref{theorem_theft} and \ref{theorem_classicical_theft}), and  it guarantees secret key expansion by securely recycling the initially pre-shared classical information (see Theorem \ref{theorem_key_recycling}). Moreover, on the infrastructure level, our scheme only requires a unidirectional insecure quantum channel from Alice to Bob, and a bidirectional insecure classical channel, aligning with the hardware requirements of the BB84 protocol.

\subsection{Drawbacks}
\label{drawbacks}

 Firstly, our scheme requires  generation and maintenance of entanglement.  A mathematical security proof accounting for imperfections of this delicate resource would need to upper-bound them ad hoc, considering near-term hardware constraints. Additionally, the  qubit string that Alice needs to send to Bob must be sent from the same location where entanglement is generated, as movable nodes cannot yet transport stored quantum systems\footnote{At present, to the best of the authors knowledge, storing and transporting quantum systems using what could be termed a 'quantum pocket' is not feasible. The recent work in \cite{q_pocket} exemplifies this by describing their proposed quantum memory as portable, although it remains non-operational during transit.}.

  Secondly, our proposal requires to, in order to guarantee authentication, double the amount of sent qubits by Alice at each round. This effectively lowers the key-generation rate by a factor of $2$.

\subsection{Benchmarking}

In the state of the art of QKD authentication \cite{qkd_authentication} and, more generally, in that of authenticated communication \cite{authenticated_communication}, there is not a  unique metric allowing for a comprehensive quantitative or qualitative comparison among the existing methods. A common trade-off in this field involves balancing security with the amount of pre-shared secrecy required, which may initially seem unavoidable.

\subsubsection{Comparison with methods not using pre-shared secrets} 

 Biometric techniques can challenge the aforementioned intuition, but concerns such as spoofing \cite{spoofing}, non-refreshability, and privacy leakage prevent them from being the preferred choice in practice. Alternatively, physical unclonable functions (PUFs) provide a token-based solution that also eliminates the need for secret pre-sharing. Notable examples include \cite{token_1}, which introduces a CPUF scheme for multi-entity authenticated QKD, and \cite{qpuf_communication}, which employs a QPUF for unidirectional communication authentication. Nevertheless, as already discussed in the introduction, PUFs rely on the critical assumption of hardware unclonability, unlike our proposed solution.

Also relevant in the state of the art, strategies relying on a trusted party represent the final approach we introduce that does not require pre-shared secrecy. A notable example can be found in the proposal \cite{experimental_PQC}, which leverages short-term security guarantees to establish unconditional security. Specifically, the cited authors propose post-quantum cryptography as a means of achieving authenticated QKD over a network, providing provable resistance against distinct types of classical- and quantum-capable adversaries, and desirably resisting replay attacks. However, and despite its effectiveness, these methods remain conjectural and may eventually become obsolete.

Except for biometrics or token-based methods, and unless a trusted entity or authenticated components are involved, all existing proposals for authenticated communication require pre-shared secrecy, as discussed both in \cite{authentication_without_secrets} and \cite{no_secrets}. Consistently, one of the most  efficient ways of message authentication today involves the use of hash functions \cite{hash}, which require pre-shared secrecy. In \cite{pseudorandom}, a classically authenticated QKD scheme is claimed to require no pre-shared secrecy by means of pseudorandomness and its efficient distinguishability from true randomness. However, they rely on an auxiliary  authenticated classical channel.

\subsubsection{Comparison with classical methods using  pre-shared secrets}

Currently, distinct and contrasted methods exist for classically authenticating QKD protocols by using pre-shared secrecy. A straightforward approach, already proposed in \cite{bennet_brassard}, is to use Message Authentication Codes (MACs) to authenticate all required communication rounds over the public classical channel. The Universal-Hashing-based  MAC (UHMAC), which offers information-theoretic security, is regarded as a reliable choice for this task, as shown in \cite{qkd_tokyo} and \cite{UMAC}. Notably, the Wegman Carter UHMAC \cite{WEGMAN_mac}, allows for hash-key recycling \cite{key_recycling}, while the One-Time Pad (OTP)  it uses to encrypt the message tag must be refreshed each round. However, the latter can be achieved by using a small portion of the  generated key, introducing key chaining and thus allowing for pre-shared secrecy amplification. Alternatively, MACs based on cryptographic hash functions (HMAC) \cite{CMAC} retain the key reusability feature and can eliminate the need for an OTP, potentially increasing the key generation rate at the cost of only providing computational security.

\begin{remark}{On the vulnerabilities of MAC-based authenticated QKD}

It is important to notice that, for the Wegman-Carter UHMAC, cryptographic primitives, such as pseudorandomness, are generally still necessary. In such cases, the security remains computational rather than information-theoretic. 
Moreover, HMACs are vulnerable to replay attacks, requiring the additional cost implied by the use of timestamps or \textit{nonces}.
\end{remark}

Alternatively, the work in  \cite{no_public_bases} proposes to authenticate prepare-and-measure QKD protocols via pre-sharing the qubit-encoding bases. This solution allows to increase the key-generation rate, but has been found to be weak against attacks that learn and accumulate partial information in each QKD round. The authors in \cite{key_update} acknowledge and prove the latter, and propose an improved method based on updating the pre-shared secrecy via universal hashing. Although they solve the weaknesses in \cite{no_public_bases}, their new protocol requires additional authentication means for an initial subset of QKD rounds.

Compared to MAC-based proposals, our entanglement-based scheme offers a solution to the QKD authentication problem that is inherently resistant to replay attacks. Furthermore, unlike the proposal in \cite{no_public_bases}, our scheme supports key recycling (see Theorem \ref{theorem_key_recycling}) and does not rely on additional authentication sources like in \cite{key_update}. Critically, and contrary to all the classical methods discussed, our scheme is resistant to the complete leakage of pre-shared classical secrets. That is, accessing non-pre-shared secrecy (owned by Bob) or quantum memories (owned by Alice) is also a necessity for a successful attack.

\subsubsection{Comparison with quantum methods using pre-shared secrets}

\begin{figure*}
\centering
\includegraphics[width=1.0\textwidth,clip]{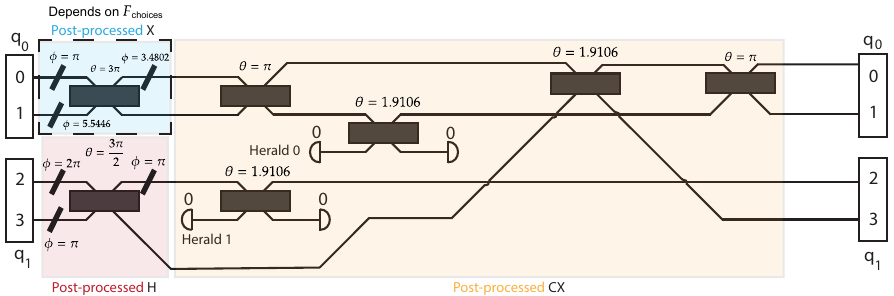}

\caption{A spatially encoded optical circuit for generating the entangled pairs required in our protocol. \(X\) or the identity operator $\mathbf{I}$ are applied according to the corresponding bit in  \(F_{\mathrm{Choices}}\).   The phase-shifter and beam-splitter parameters, \(\phi\) and \(\theta\), respectively, control the optical transformations within the circuit.}
\label{Fig:optical_circ}    
\end{figure*}

On the one hand, an initial proposal for entanglement-assisted authenticated quantum key distribution (QKD) is presented in \cite{entanglement_assisted_qkd}. In this work, the authors introduce a protocol that leverages maximally entangled pairs and that, as our scheme, requires quantum memories. While their protocol eliminates the need for classical communication, it remains susceptible to a straightforward man-in-the-middle attack due to the lack of pre-shared secrecy.

 On the other hand, while our work may  resemble the ideas from \cite{main_citation}, we present a different authenticated QKD procedure, with distinct hardware requirements and derived theoretical properties. First, the cited article does not address the possibility of recycling the pre-shared key used in its scheme, or of establishing key chaining across successive sessions. Second, while the cited proposal conveniently does not rely on entanglement generation and maintenance, it can be broken by the mere leakage of pre-shared classical information\footnote{The cited authors encode pre-shared classical information into quantum systems. However, in their scheme, if an attacker is aware of the encoding and decoding rules and gains access to the pre-shared classical information, they can impersonate either party involved without detection.}.

\section{Simulation}

    \begin{table*}
\centering
\caption{Simulation parameters}
\begin{tabular}{ |p{4.5cm}|p{2cm}|p{1cm}|p{4.5cm}|p{2cm}|p{1cm}| }
\hline
\textbf{Parameter} & \textbf{Value} & \textbf{Unit} & \textbf{Parameter} & \textbf{Value} & \textbf{Unit}\\
\hline
Qubit decoherence time ($T_{1}$) & $2.2344 \cdot 10^2$ & $\mu \mathrm{s}$ & Comb finesse ($F$) & $4\cdot 10$ & -\\
\hline
Qubit dephasing time ($T_{2}$) & $2.9535\cdot 10^2$ & $\mu \mathrm{s}$ & Comb absorption efficiency ($\alpha l$) & $1.0$ &  - \\
\hline
Reflectivity mirror 1 ($R_{1}$) & $9.6\cdot 10^{-1}$ & - & Comb FWHM linewidth ($\epsilon$)& $3.0$ & $\mathrm{kHz}$ \\
\hline
 Reflectivity mirror 2 ($R_{2}$) & $9.9\cdot 10^{-1}$ &  -  & Number of photons Fig.\,\ref{proto:fig}/ Fig.\,\ref{Fig:prob} ($2\lambda$) & $10^3$/ $2\cdot 10^2$ &  - \\
\hline
Source frequency ($f_{\mathrm{source}}$)  & $3.3\cdot 10$ & $\mathrm{MHz}$  
 & Detection efficiency ($p_{\mathrm{detect}}$)  & $9.5 \cdot 10^{-1}$ & -  \\
\hline
Source wavelength & $1.550\cdot 10^3$& $\mathrm{nm}$ 
 & Fiber attenuation ($\tau$) & $1.7\cdot10^{-1}$ & $\mathrm{dBkm}^{-1}$  \\
\hline
Driven recovery time & $3.0\cdot10$& $\mathrm{ns}$
& First lens brightness of quantum dot & $9.0\cdot10^{-1}$& - \\
\hline
Driven storage time & $3.0\cdot 10$ & $\mathrm{ns}$
 &$g^{(2)}$ of source & $1.0\cdot10^{-2}$ & - \\
\hline
 Photon velocity in fiber& $2.08\cdot10^{8}$ & $\mathrm{ms}^{-1}$
 & Photon distinguishability & $9.5\cdot10^{-1}$ &  - \\
\hline
  Dark-count frequency ($f_{\mathrm{dark}}$)  & $1.0\cdot 10$ & $\mathrm{Hz}$ & Average BS and PS operation time & $1.0\cdot10^{1}$ &  $\mathrm{ns}$\\
\hline
\end{tabular}
\label{parameter:table} 
\end{table*}  

\subsection{Methods}
\label{realistic}
In this section, we outline the methodology employed to simulate our noise-adapted QKD authentication protocol in a realistic scenario. The parameters characterizing our model are detailed in Table.\,\ref{parameter:table}, and all simulations described in this paper were parallelized \cite{multiprocessing} on a 2x Intel Xeon Platinum 8176 @ 2.1 GHz processor.

\subsubsection{Simulation processing}
The simulation adopts a shot-by-shot (photon-by-photon) approach, where a scheduler is implemented to monitor the local time of each photon at every stage of the simulation. 
This method, together with a unification of simulation platforms via process tomography, enables to estimate the quantum channel undergone by the involved quantum systems between entanglement creation and authentication stages. 

 At each step of the simulation, we update the density matrix $\rho$ corresponding to the $2$-qubit system at issue according to the Chi matrix representation of a channel $\Lambda$, defined as
\begin{equation}
\Lambda(\rho)=\sum_{i,j= 0}^{15} \chi_{i,j}P_{i}\rho P_{j}.
\label{Chi_m}
\end{equation}
Here,  $\chi_{i,j}$ defines the matrix representation of  $\Lambda$,  and $P_x$ the $x$-th $2$-qubit Pauli-basis operator. These Chi matrices are derived through process tomography, where \(10^3\) systems are  initialized in different states building a tomographically complete set.

\subsubsection{Entanglement creation}

We modeled the source as an imperfect quantum-dot-based single-photon emitter, represented by a mixture of Fock states, as described in \cite{A2}. The source has a second-order intensity autocorrelation at zero time delay, \( g^{(2)} = 0.01 \), a two-photon mean wavepacket overlap (photon indistinguishability) of 0.95, and a first-lens brightness of 0.9. The source operates at a frequency of \( f_{\mathrm{source}} = 33 \) MHz, as reported in \cite{Liu2021}. The wavelength of the source was specifically selected to be compatible with AFC quantum memories, avoiding additional frequency conversion losses, as demonstrated with erbium-doped crystals in \cite{Rančić2018}.

We conducted an optical simulation using a quantum optical processor at both communication ends, implemented with the Perceval package and its \textit{Naive} backend \cite{Heurtel_2023}. To account for imperfections, we modified the package to include a stochastic photon loss model for the beam splitters and phase shifters, following a random walk-based approach as described in \cite{PhysRevResearch.6.033337}.

In this model, the photon loss of the beam splitter (BS) is described by:

\begin{equation}
    BS_{\mathrm{loss}}=\begin{bmatrix}
\mathrm{cos}(\frac{\theta}{2})A_{01} & i\mathrm{sin}(\frac{\theta}{2})B_{01} \\
i\mathrm{sin}(\frac{\theta}{2})A_{01} & \mathrm{cos}(\frac{\theta}{2})B_{01}
\end{bmatrix},
\end{equation}
where \( A_{01} = \cos(\epsilon_{0} I_{\mathrm{c}_0}) \cos(\epsilon_{1} I_{\mathrm{s}_1}) \) and \( B_{01} = \cos(\epsilon_{0} I_{\mathrm{s}_0}) \cos(\epsilon_{1} I_{\mathrm{c}_1}) \). Here, \( I_{c_i} \) and \( I_{s_i} \) are defined as follows:

\[
I_{c_i} = \int_{0}^{t} \cos(\theta(0,s)) \, dW_{i}(s),
\]

\[
I_{s_i} = \int_{0}^{t} \sin(\theta(0,s)) \, dW_{i}(s).
\]
These correspond to Itô stochastic integrals which sum the averages of a Wiener process \( W \) over the time interval \( \Delta t \). The parameter \( \epsilon_{i} \) is defined as:

\[
\epsilon_{i} = \sqrt{\frac{-\log(1-2p_{i})}{2}},
\]
where \( p_{i} = 1 - e^{-2\frac{\Delta t}{T_{1}}} \) represents the photon loss probability of mode \( i \) in the optical channel. In this context, \( \Delta t \approx 10 \, \text{ns} \) is the gate time of the component, and \( T_{1} \approx 2.23 \times 10^{2} \, \mu \mathrm{s} \) is the photon decoherence time.

In Fig.\,\ref{Fig:optical_circ}, we illustrate the optical setup simulated for the entanglement generation stage within the  authenticated QKD protocol. The entanglement creation process is encoded across four spatial modes, with two auxiliary heralding modes for $CX$ gate generation.

\subsubsection{Channel attenuation}
\label{channel}

The fiber loss channel is characterized by the single-photon transmission probability \cite{Coopmans2021}:

\begin{equation}
    \eta_{\mathrm{channel}} = 10^{-d\tau/10},
    \label{chan}
\end{equation}
where \(d\) represents the transmission distance between parties, and \(\tau = 0.17 \,\mathrm{dBkm}^{-1}\) denotes the fiber attenuation. The refractive index of the glass in the fiber is \(r_{\mathrm{glass}} = 1.44\).

Prior to measurement, photons are detected from the optical fiber  channel. We assume a Superconducting Nanowire Single-Photon
Detector (SNSPD) with a detection efficiency of \(p_{\mathrm{detect}} = 0.9\) \cite{Liu2021}.

The dark-count probability is governed by a Poisson distribution \cite{Coopmans2021}:
\begin{equation}
    \eta_{\mathrm{dark}} = 1 - e^{-t_{\mathrm{w}} f_{\mathrm{dark}}},
    \label{chan2}
\end{equation}
where the detector's capture window is \(t_{\mathrm{w}} = 25 \,\mathrm{ns}\) \cite{supplementary_window}, and \(f_{\mathrm{dark}}\) is the dark-count frequency. Experimentally, \(f_{\mathrm{dark}}\) has been determined to be approximately \(10\,\mathrm{Hz}\) \cite{Liu2021}, a value used in our simulations.

%, in \cite{Liu2021}  %$t_{\mathrm{w}}=5\,\mathrm{ns}
%$ is considered for the AFC. 

\subsubsection{Quantum memory}
\label{Memm}

At Alice's side, we consider multiple AFC memories \cite{duranti2023efficient}, which directly couple to the photon's radio frequency without requiring additional conversions, as is necessary with nitrogen-vacancy centers or transmon-coupled cavities. State-of-the-art single-mode AFC storage has demonstrated a maximum coherent storage time of up to one hour \cite{Ma2021} by employing a zero-first-order Zeeman magnetic field and dynamical decoupling to preserve spin coherence, achieving a fidelity of 96.4\%. 

For multimode AFC memories, significant advancements have been achieved, including the demonstration of over 15 spatial × 30 temporal modes \cite{PhysRevLett.123.080502} and 1060 temporal modes \cite{Bonarota_2011}. Storage lifetimes for multimode systems have reached \(0.542 \,\mathrm{ms}\) \cite{PhysRevA.93.032327}. Additionally, the dephasing time \(T_2\) for AFC has been experimentally measured to be \(T_{2} = (300 \pm 30)\,\mu\mathrm{s}\) \cite{Ortu_2022}.

More specifically, we are considering Stark-modulated AFC memories with an optical cavity \cite{PhysRevResearch.3.023099} and the following considerations:

\begin{figure*}

\begin{minipage}[t]{0.495\textwidth}
    \includegraphics[width=1.0\linewidth]{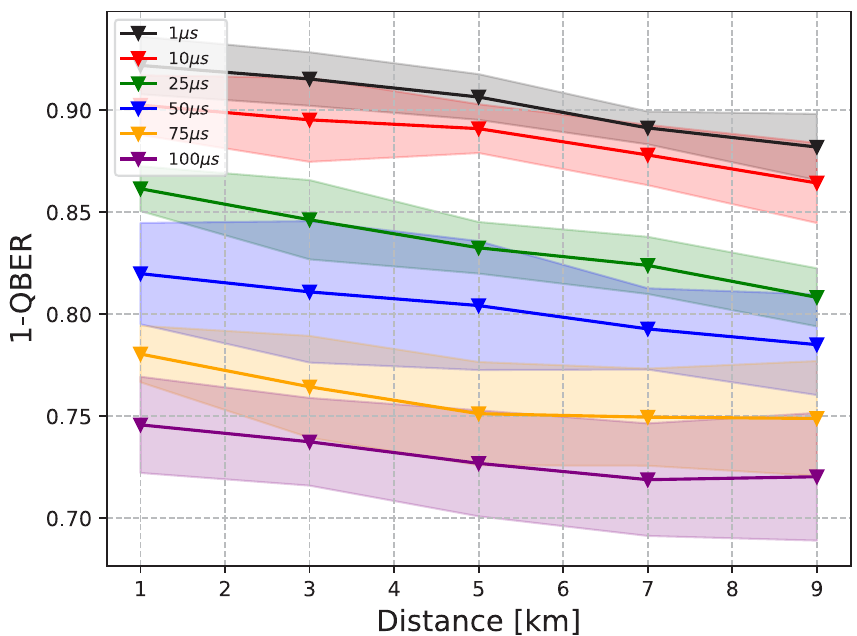}
  \end{minipage}
  \hfill
  \begin{minipage}[t]{0.495\textwidth}
    \includegraphics[width=1.1\linewidth]{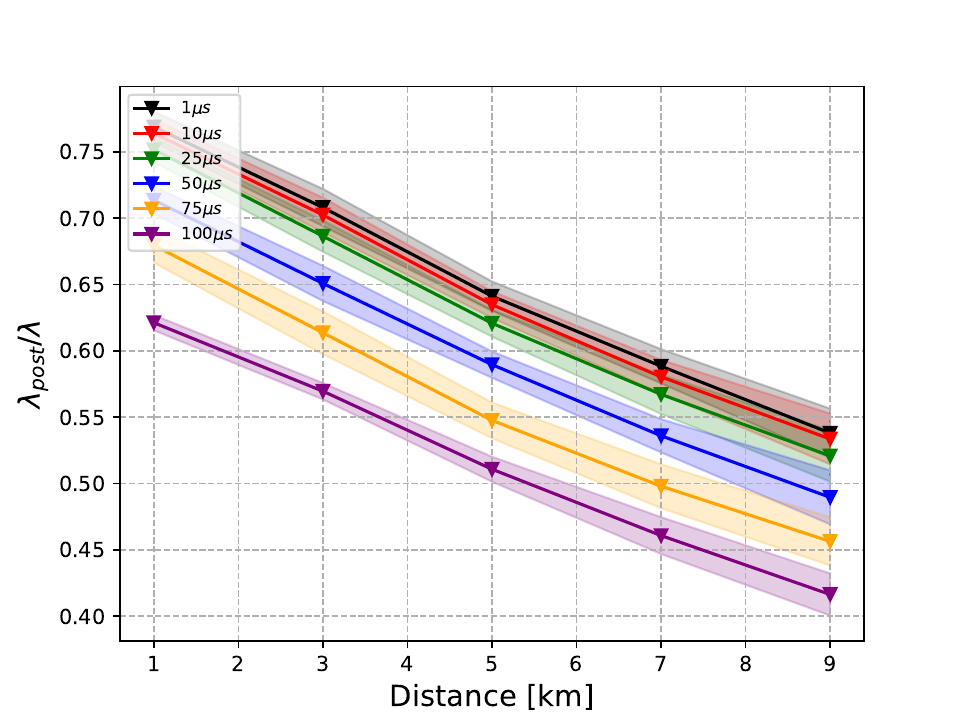}
  \end{minipage}
  
  \caption{BB84 authentication performance, with the simulation parameters presented in Table.\,\ref{parameter:table} and $\lambda = 500$: \textbf{Left)} $1-\text{QBER}$ against transmission distance for multiple storage times. \textbf{Right)} Ratio  $\lambda_{\mathrm{post}}/\lambda$ against distance for multiple storage times. Five samples, each generated with a different seed, were used to calculate the standard deviation.}
\label{proto:fig}
\end{figure*}

\begin{itemize}
    \item We use as many cavity-enhanced AFC memories as needed for the size of the quantum key.

    \item We set\footnote{Such value, not currently feasible (see Table.\,\ref{parameters:state}), has been set ad hoc to increase the memory retrieval efficiency,  in order to obtain informational simulations.} the finesse of the comb $F_{\mathrm{AFC}}=40$, which relates with the memory retrieval probability   $\eta_{\mathrm{cav}} $ as\cite{PhysRevResearch.3.023099}:

    \begin{equation}
    \eta_{\mathrm{cav}}=\frac{4(\bar\alpha l)^{2}e^{-2\bar\alpha l}(1-R_{1})^{2}R_{2}e^{-t^{2}\bar\epsilon^{2}}}{(1-\sqrt{R_{1}R_{2}}e^{-\bar\alpha l})^{4}},
    \label{eq: cav}
\end{equation}
where $t$ is the storage time of the photon, $\bar\alpha= \frac{\alpha}{F_{\mathrm{AFC}}}\sqrt{\frac{\pi}{4\mathrm{ln(2)}}}$ is the effective absorption of the comb, $\alpha$ is the absorption coefficient of the comb peaks and $l$ is the crystal length. $R_{1}=0.96$ and $R_{2}=0.99$ are the mirror reflectivities and $\bar\epsilon=\frac{2\pi\epsilon}{\sqrt{8\mathrm{ln(2)}}}$ is related to the comb FWHM $\epsilon$.

\end{itemize}

The storage and retrieval efficiency involves a trade-off when choosing $F_{\mathrm{AFC}}$: higher absorption probability corresponds to a smaller $F_{\mathrm{AFC}}$, while increasing $F_{\mathrm{AFC}}$ reduces dephasing during storage \cite{PhysRevA.79.052329}. In our simulation we have opted for $F=40$ and $\alpha l = 1$, with $\epsilon$ of $3 \,\mathrm{kHz}$ and with negligible inter-cavity loss. Experimentally, with $F_{\mathrm{AFC}}=5.8$, storage-retrieval efficiencies of $(55\pm5)\%$ \cite{PhysRevLett.110.133604} and $62\%$ \cite{duranti2023efficient} were obtained for cavity coupled AFCs, but we expect that the parameters chosen for our simulation may be achievable for AFC with persistent holes and with $\epsilon$ in the order of $\mathrm{kHz}$ in the near future \cite{PhysRevResearch.3.023099}. Despite of its relevance, in the simulation we have not considered the dependence of the added atomic dephasing due to the change in comb finesse \cite{PhysRevA.82.022310}. In Table.\,\ref{parameters:state}, we provide an overview of the main parameters from cavity enhanced AFC experiments developed through the past  years. 

For the simulation, we have considered no added losses when converting from $\mathrm{kHz}$ to telecom frequency, which may be achievable for single photons \cite{MA201269}. Additionally, we have assumed that the post-conversion fidelity remains unchanged, and we have considered driven storage and recovery times of $30 \,\mathrm{ns}$ for each photon.

\subsubsection{DNN post-processing}
\label{post}
We propose to evaluate the post-processing performance in a noisy environment by simulating an impersonation attack on Bob. During the authentication phase, the attacker is assumed to access the sent qubit string while lacking knowledge of \(K_{\text{Positions}}\).

To adapt our QKD authentication protocol to noise, we have first introduced the parameter $\mu$, which, under low noise assumptions and by setting it to be small enough, ensures that legitimate entities are accepted with high probability. The trade-off on $\mu$ is set by the fact that it needs to be large in order to ensure rejection on forgery attempts, as suggested by  Eq.\,\eqref{mu_equation}.

Alternatively, we address the problem of distinguishing legitimate parties from attackers as a binary classification task utilizing a Deep Neural Network (DNN). This approach focuses on differentiating attackers from legitimate entities by analyzing the authentication qubit measurements for both parties. To this end, we generate a dataset consisting of measurements from \( 6.5 \cdot 10^3 \) valid parties and \( 6.5 \cdot 10^3 \) attackers, with  \( \lambda = 100 \). The dataset is partitioned, with $90\%$ allocated for training and the remainder reserved for validation.

The DNN architecture used in this study consists of two hidden layers with 20 and 10 nodes, respectively. Each layer employs the Rectified Linear Unit (ReLU) activation function, while the output layer uses the sigmoid activation function. The model is trained using the binary cross-entropy loss function and the Adam optimizer. To determine the optimal hyperparameters (depth and width) for the DNN given a specific \( \lambda \), we evaluate five different arbitrary configurations of the DNN architecture and select the one that achieves the best performance.

\subsection{Results}
\label{results}

In this section, we start showing the noise impact on our derived protocol. Subsequently, we evaluate its performance in a more detailed basis, comparing the performance of the two proposed noise adaptations.

\subsubsection{Noise impact}
The authentication protocol performance is illustrated in Fig.\,\ref{proto:fig} and was simulated for various distances \( d \), spanning \( d \in [1, 10] \, \mathrm{km} \). After generating \( 2\lambda =   10^3 \) photons, both parties wait for a duration \( T \in [1, 100] \, \mu \mathrm{s} \) before Alice transmits the qubits.

For each data point in Fig.\,\ref{proto:fig} (left), the mean and variance of $1-\text{QBER}$ were calculated by simulating $5$ different and independent sessions. The high variance values are attributed to the low number of AU photonic qubits (\( \lambda = 500 \)) used. This photon count was intentionally kept low to minimize storage and processing times, as increasing the number of photons would significantly extend both. 

In Fig.\,\ref{proto:fig} (right), we analyze photon losses during transmission and storage by examining the ratio \((\lambda_{\text{post}} / \lambda)\), where \(\lambda_{\text{post}}\) represents the number of transmitted photons available after losses, and \(\lambda\) is the number of AU photons initially sent. As shown in Fig.\,\ref{proto:fig} (right), the ratio decreases exponentially with increasing transmission distance, following Eq.\,\eqref{chan} and Eq.\,\eqref{chan2}, and with longer storage times due to the storage and retrieval efficiencies of the AFC memory, as described by Eq.\,\eqref{eq: cav}. 

\begin{figure}
\includegraphics[width=0.48\textwidth,clip]{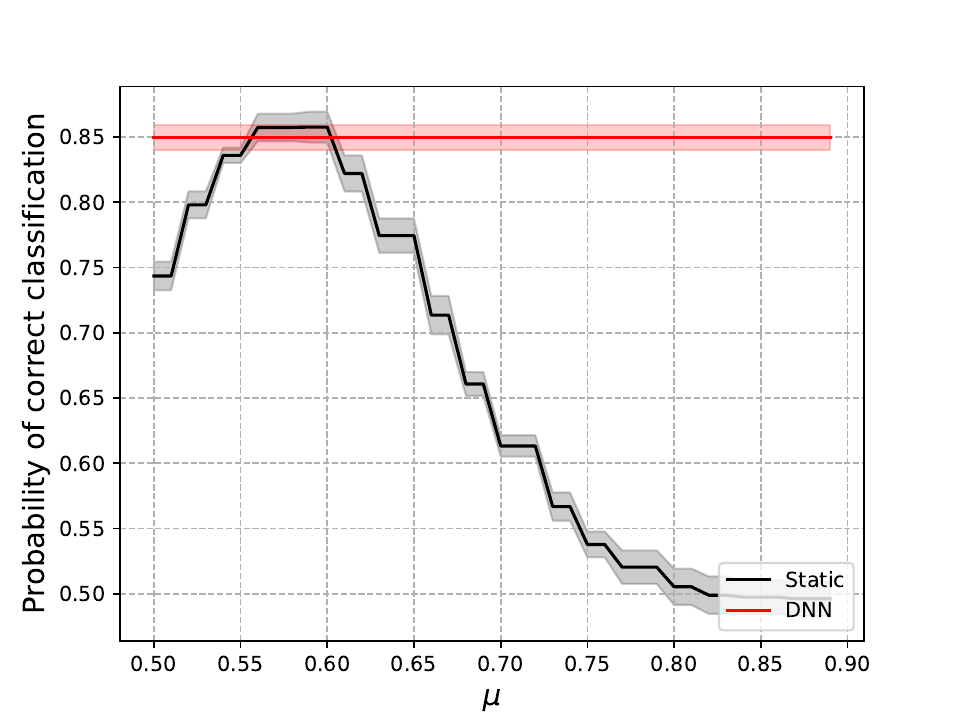}
\caption{Comparison, between the static and the DNN methods, of the relative frequency  obtained for correctly distinguishing between valid entities and attackers, for \(\lambda = 100\). We set a distance between parties of $d= 1$ km, a memory storage time of $ T = 150$ $\mu\mathrm{s}$ and we calculate  the standard deviation  from 10 different simulations, each with distinct training and test set shuffling.}

\label{Fig:prob}       
\end{figure}

\subsubsection{Post-processing DNN discrimination performance}

In Fig.\,\ref{Fig:prob}, we compare two methods for the binary classification problem of distinguishing attackers from legitimate parties using either the acceptance condition a) or b) (see Section\,\ref{Noise_adaptation}).

For a distance of \(1\) $\mathrm{km}$ and a memory storage time of $150$ $\mu s$, we sample a range of values of \( \mu \) to identify the optimal bound and compare this peak accuracy to that of the DNN. For an authentication protocol with \( \lambda = 100 \), both the static and DNN methods achieve comparable accuracy, exceeding \( 0.80 \). 

Using the SHAP library \cite{shap}, we have verified the impact of the features on the model, observing that noise has a homogeneous effect across the entire set of photons. This prevents data from valid entities from exhibiting a more complex structure, which we conjecture would make the DNN method preferable.

\section{Conclusions}
\label{conclusions}

In this article, we begin by introducing the key concepts and terms essential for understanding our research. We then review the paradigmatic BB84 protocol, emphasizing its main  steps and features. Following this, we present an entanglement-assisted method for authenticating such protocol. In the noiseless scenario, and under two simple assumptions on pre-shared and non-pre-shared information, and access to quantum memories, our proposal offers  information-theoretically secure authentication. To account for noisy conditions, we adapt our scheme using both a statistical approach and a DNN-aided heuristic method.

We then compare our noiseless derivations with the current state of the art in QKD authentication. Specifically, we highlight how our protocol resists the leakage of all pre-shared secrets by utilizing non-pre-shared secrecy through entanglement. This sets it apart from existing quantum and classical methods which rely on pre-shared secrecy. Notably, our approach enables the recycling of the single pre-shared key required across multiple sessions, making our proposal preferable  to other existing quantum methods and aligning it with classical schemes in this regard.

Within a numerical simulation, our  protocol is evaluated under noisy channel conditions utilizing quantum optical processors and AFC cavity-enhanced memories, spanning distances ranging from 1 to 10 $\mathrm{km}$ and storage times up to 100 $\mu \mathrm{s}$. Feasibility requirements are established, observing, for a 1 km distance and a storage time of 10 $\mu \mathrm{s}$, a  $\text{QBER}$ of 0.1. Moreover,  both a static method and a deep neural network approach are proposed in order to distinguish between legitimate parties and forgery attempts. Both achieve a correct classification rate larger than 0.80 for memory storage time of 150 $\mu \mathrm{s}$ and a 1 $\mathrm{km}$ distance between  communication parties.

Before concluding this section, we aim to stress that creating a quantum-based authentication protocol with provable security remains an open challenge, unless bounds on the amount of noise are  set. In this line, we reckon that  our realistic simulation openly tackles the current scope of our noiseless derivation, together with providing the reader with deep insights on the state of the art of AFC quantum memories. 

\section{Further research}
\label{further_research}
We foresee several avenues for future research. Specifically, we contend that our contributions may serve as a foundation for the development of future protocols. One example of a  line of research, within the framework of quantum authenticated QKD, could be  exploiting already existing quantum models for MACs, e.g., the one presented in \cite{qmac}. Moreover, it would be of great interest to generalize our protocol towards a quantum network, finding scaling laws as well as adequate comparison metrics with the state of the art in the field. 

With respect to the noise-adapted protocol versions, our Deep-Neural-Network (DNN)-based scheme does not clearly represent an advantage with respect to a our other, simpler, proposal. Moreover,  we only prove soundness against a randomized attacker. Therefore, we  propose to both leverage induced or inferred  noise structure in order to boost the DNN-method performance, and  to investigate more complex adversaries. Additionally, in the context of non-uniformity of noise, we suggest exploring alternative algorithms, e.g.,  convolutional neural networks \cite{Yamashita2018} intended to  patterns recognition in extensive matrix input data, as shown for image recognition. 

Finally, as another direction for future research, we propose to investigate the plausibility of the assumptions that underpin the security of our protocol (see Section \ref{sec analysis}). Violations of these assumptions could be tackled by carefully scheduling and monitoring the time slots allocated for interactions between Alice and Bob, thereby adding complexity to the protocol. If any party publicly discloses $K_{\mathrm{Positions}}$ while claiming exclusion from a specific stage, the protocol should be aborted.

\section*{Code Availability}
\label{code}
All codes responsible for the results in this article can be found at \cite{code_repo}.

\section*{Acknowledgements}
The authors acknowledge valuable discussions with Jonas Hawellek, Pin-Hsun Lin, Dagmar Bruß and Hermann Kampermann, as well as the reviewers of the journal Physica Scripta for their accurate and constructive comments and suggestions. The authors acknowledge the financial support by the Federal Ministry of Education and Research of Germany in the program of “Souverän. Digital. Vernetzt”. Joint project 6G-life, project identification numbers: 16KISK002 and 16KISK263. Furthermore, VG and JN acknowledge the financial support under projects 16KISQ039, 16KISQ077 and 16KISQ168 and by the DFG via project NO 1129/2-1. CD and PJF were further supported under projects 16KISQ169, 16KIS2196, 16KISQ038, 16KISR038, 16KISQ0170, and 16KIS2234.

\appendix
\renewcommand{\thesection}{\Alph{section}}

\subsection{Propositions and theorems}
\label{appendix_proofs}

\begin{proposition}
\label{proposition_completeness}
The  protocol described in Section \ref{noiseless protocol} is $0$-complete under noiseless conditions.
\end{proposition}

In order to prove this proposition,  we show that in the absence of noise, and provided that the quantum systems involved are not tampered with, the aforementioned authenticated QKD  protocol  is not aborted.

\begin{proof}
    At any given round, an entry of \( F_{\mathrm{Choices}} \) takes the value "\( 0 \)" whenever \( \mathbf{I} \) is the chosen gate applied in the circuit shown in Fig.\,\ref{circuit}. In this case, the corresponding two-qubit state of the control-target bipartite system, just before measurement, is

\begin{equation}
        \ket{\psi_{\mathrm{pre-meas}}}_{\mathrm{CT}_{\mathrm{"0"}}} =  \frac{1}{\sqrt{2}}\big( \ket{0}_{\mathrm{C}} \otimes \ket{0}_{\mathrm{T}} + \ket{1}_{\mathrm{C}} \otimes \ket{1}_{\mathrm{T}}\big).
\end{equation}
Thus, performing the XOR (\(\oplus\)) operation on the two measurement outcomes, obtained by measuring each qubit in the computational basis, consistently yields a "\(0\)". This ensures that Alice and Bob correctly exchange the basis choice associated with the considered entry in \( F_{\mathrm{Choices}} \) (see Eqs.~\eqref{ecrypt1}, \eqref{assume1}, \eqref{ecrypt2}, and \eqref{assume2}).

On the other hand, when \( X \) is chosen instead of \( \mathbf{I} \), the entries of \( F_{\mathrm{Choices}} \) take the value "\( 1 \)". In this case, the corresponding generated two-qubit state takes the form

\begin{equation}
        \ket{\psi_{\mathrm{pre-meas}}}_{\mathrm{CT}_{\mathrm{"1"}}} =  \frac{1}{\sqrt{2}}\big( \ket{0}_{\mathrm{C}} \otimes \ket{1}_{\mathrm{T}} + \ket{1}_{\mathrm{C}} \otimes \ket{0}_{\mathrm{T}}\big).
\end{equation}
Thus, performing the XOR (\(\oplus\)) operation on the two measurement outcomes, obtained by measuring each qubit in the computational basis, consistently yields a "\(1\)". This ensures that Alice and Bob correctly exchange the basis choice associated with the considered entry in \( F_{\mathrm{Choices}} \) (see Eqs.~\eqref{ecrypt1}, \eqref{assume1}, \eqref{ecrypt2}, and \eqref{assume2}).

\end{proof}

\begin{theorem}
    The  protocol described in Section \ref{noiseless protocol} is $\big(0, \big(\frac{3}{4}\big)^{\xi\lambda}\big)$-secure under noiseless conditions. \label{theorem_security}
\end{theorem}
\begin{proof}
In order to prove security, we impose the two assumptions made in Section \ref{sec analysis}, while sticking to Definition \ref{security_def}. In this line, and given that Proposition \ref{proposition_completeness} asserts completeness of our scheme, we aim to prove soundness, in the sense described in Definition \ref{soundness_def}. Theorems \ref{theorem_theft} and \ref{theorem_classicical_theft} presented ensure one-session security against adversaries that exhaust all the adversarial resources allowed by our model. Finally, Theorem \ref{theorem_key_recycling} states that multiple sessions can be run without the need of refreshing the pre-shared key $K_{\mathrm{Positions}}$.

\begin{subtheorem}
    Under noiseless conditions, the protocol described in Section \ref{noiseless protocol} is $\Big(0, \frac{1}{2^{\xi \lambda}}   \Big)$-secure against adversaries having access to the pre-shared key $K_{\mathrm{Positions}}$ if they neither have access to the non-shared key $F_{\mathrm{Choices}}$, owned by Bob, nor to the quantum memories, held by Alice. 
    \label{theorem_theft}
\end{subtheorem}

\begin{proof}

We will firstly (secondly) prove security against adversaries aiming to forge Alice (Bob):

\begin{enumerate}[label=\alph*.]

\item We model the maximum capabilities of an adversary that owns $K_{\mathrm{Positions}}$ and tries  to forge Alice's identity by allowing such adversary to prepare $\lambda$ maximally entangled states, whose exact preparation is defined by a bitstring $F_{\mathrm{Choices\text{-}\mathrm{Adv}}}$. Subsequently, the adversary proceeds with the protocol, starting from the 5th step, and substituting $F_{\mathrm{Alice}}$ by the outcomes obtained on the alternatively prepared states.

Because $F_{\mathrm{Choices\text{-}\mathrm{Adv}}}$ and $F_{\mathrm{Choices}}$ are not correlated, this strategy completely randomizes  the encrypted bases exchange, leading to an expected QBER of $0.5$ within the fraction $\xi$ checked of the sifted key. Specifically, a null QBER, needed for a successful forgery, is obtained with a probability equal to $\frac{1}{2^{\xi \lambda}}$.

\item We model the maximum capabilities of an adversary  that owns  $K_{\mathrm{Positions}}$ and tries  to forge Bob's identity by allowing such adversary to perform a straightforward man-in-the-middle attack:  the adversary  intercepts all qubits and measures them in the computational basis so that $F_{\mathrm{Bob}}$ is obtained. 

The adversary, nonetheless, still needs to make a random guess about $F_{\mathrm{Choices}}$. This, again, completely randomizes the encrypted bases exchange, again leading to an expected QBER of $0.5$ within the compared fraction $\xi$. Specifically, a null QBER, needed for a successful forgery, is obtained with a  probability equal to $\frac{1}{2^{\xi \lambda}}$.
\end{enumerate}

\end{proof}

\begin{table*}
\centering
\caption{Hardware parameters: values overview}
\begin{tabular}{ |p{4.5cm}|p{2.5cm}|p{2cm}|p{3.5cm}|p{3cm}| }
\hline
\textbf{Previous works} & \textbf{Comb finesse ($F$)} & \textbf{Storage time} & \textbf{Absorption coefficient ($\alpha$l)} & \textbf{Storage efficiency}\\
\hline

Mikael Afzelius. et al. \cite{AMARI20101579} 2010 & $2.5$ & $20$ $\mu \mathrm{s}$ & $0.5$ & $1\%$\\
 \hline
 
Mahmood Sabooni et al. \cite{PhysRevLett.110.133604} 2013 & $3.0$ & $1.1$ $\mu \mathrm{s}$ & $1$ & $(58\pm5)\%$\\
\hline
 
 P Jobez et al. \cite{Jobez_2014} 2014 & $5.0$ & $2$ $\mu s$ / $10$ $\mu \mathrm{s}$ & $1.2$ & $53\% 
 / 28\%$\\
 \hline

 Jacob H. D. et al. \cite{PhysRevA.101.042333} 2020 & $2.0$ & $25$ $\mathrm{ns}$ & $0.45$ & $27.5\%$\\
 \hline
 
  Yu Ma et al. \cite{Ma2021} 2021 & $2.2$ & $60$ $\mathrm{min}$ & $2.6$ & $0.06\%$ for $5$ min storage\\
 \hline

 Stefano Duranti. et al. \cite{duranti2023efficient} 2023 & $5.8$ & $2$ $\mu \mathrm{s}$ & $0.46$ & $62\%$\\
 \hline
\end{tabular}
\label{parameters:state} 
\end{table*}

\begin{subtheorem}
\label{including_intercept_theorem}
    Under noiseless conditions, the  QKD authentication protocol described in Section \ref{noiseless protocol} is $\big(0, \big(\frac{3}{4}\big)^{\xi\lambda}\big)$-secure against adversaries having access  to the non-shared key $F_{\mathrm{Choices}}$, owned by Bob, and to the quantum memories, held by Alice, but not having access to $K_{\mathrm{Positions}}$.
    \label{theorem_classicical_theft}
\end{subtheorem}

\begin{proof}
We will firstly (secondly) prove security against adversaries aiming to forge Alice (Bob):

\begin{enumerate}[label=\alph*.]
    \item We model the maximum capabilities of an adversary  that owns $F_{\mathrm{Choices}}$, has access to the quantum memories and aims to forge Alice's identity by allowing such adversary to either steal the $\lambda$ entangled qubits at Alice's side or to prepare an identical set of entangled states by using $F_{\mathrm{Choices}}$. The adversary further measures such entangled qubits, so that $F_{\mathrm{Bob}}$ and $F_{\mathrm{Alice}}$  can also be obtained. Then, the adversary proceeds with the protocol, starting from the 5th step, but substituting $K_{\mathrm{Positions}}$ by another and uncorrelated integer string $K_{\mathrm{Positions}\text{-}\text{Adv}}$.

    The adversary inlays the $\lambda$  AU qubits among the $\lambda$ ones devoted to QKD in one of the

\begin{equation}
    S_{\mathrm{Total}} \equiv \binom{2 \lambda}{\lambda} = \frac{(2\lambda)!}{\big(\lambda!\big)^2}
    \label{total}
\end{equation}
possible ways of doing  it, uniformly at random.

Given the vector $\Vec{x}\equiv$ $x_1$, $x_2$, ..., $x_{\lambda}$ defining the guess of the adversary on what positions the list of sorted qubits devoted to QKD occupy, the expected number $\mathbb{E}[g]$ of correct guesses $g$ takes the following form:

\begin{equation}
    \mathbb{E}[g](\Vec{x}) = \sum_{i, j=0, 1}^{\lambda}
p(j, i + j) \delta_{x_j, i + j},
\label{correct}
\end{equation}
where $p(j, i +j)$ is equal to the probability of the $j$-th qubit devoted to QKD to fall into the position $i + j$. Notice that

\begin{equation}
    \max_{\substack{0\leq i \leq \lambda  \\ 1\leq j \leq \lambda}} \big\{ p(j, i + j)    \big\} = p(1, 1)  = p(\lambda, 2\lambda).
    \label{bound_1_attempt}
\end{equation}
This can be understood by the fact that a non-extreme position of a qubit devoted to QKD being fixed constraints the $\lambda-1$ remaining ones to coexist in two different segments, thus diminishing the total number of compatible configurations.

Now, notice that the number $S_{1, 1}$ of configurations that are compatible with the overall constraints and which own the  first qubit devoted to QKD at the first position is 

\begin{equation}
    S_{1, 1} = S_{\lambda, 2\lambda} \equiv \binom{2\lambda - 1}{\lambda - 1} = \frac{(2\lambda - 1)!}{(\lambda - 1)! \, \lambda!},
    \label{partial}
\end{equation}
where $S_{\lambda, 2\lambda}$ is an analogous quantity corresponding to the number  of configurations that are compatible with the overall constraints and which own the  last qubit devoted to QKD at the last position.

All together,  Eqs.\,\eqref{correct}, \eqref{bound_1_attempt} and \eqref{partial}  allow us to write

\begin{equation}
    \mathbb{E}[g](\Vec{x}) \leq \lambda \cdot p(1, 1) = \lambda  \frac{S_{1, 1}}{S_{\mathrm{Total}}} = \frac{\lambda}{2}.
\end{equation}

Finally, we observe that the expected QBER for the fraction $\xi$ compared within the sifted key is

\begin{equation}
    \mathbb{E}[QBER] = \Big(1-\frac{1}{\lambda}\mathbb{E}[g](\Vec{x}) \Big) \frac{1}{2} \geq \frac{1}{4}.
\end{equation}
Specifically,  given Eqs. \eqref{total}, \eqref{bound_1_attempt}  and \eqref{partial}, a null QBER, needed for a successful forgery, is obtained with a probability upper-bounded by the  quantity $\big(\frac{3}{4}\big)^{\xi \lambda}$.
    
\item  We model the maximum capabilities of an adversary  that owns  $F_{\mathrm{Choices}}$, has access to the quantum memories at Alice's side and tries  to forge Bob's identity by allowing such adversary  to measure the entangled states owned by Alice before she sends them,  so that $F_{\mathrm{Bob}}$ and $F_{\mathrm{Alice}}$ can be obtained. The adversary further performs a straightforward man-in-the-middle attack to  proceed with the protocol, starting from the 5th step, but substituting $K_{\mathrm{Positions}}$ by another and uncorrelated integer string $K_{\mathrm{Positions}\text{-}\text{Adv}}$.

For this part of the proof, we can built an analogous argument as the one built against adversaries aiming to forge Alice's identity.
That is, not knowing $K_{\mathrm{Positions}}$ will again randomize, for the adversary, the position of those qubits devoted to QKD, again yielding an expected QBER for the sifted key lower-bounded by  $\frac{1}{4}$. Specifically, given Eqs. \eqref{total}, \eqref{bound_1_attempt} and \eqref{partial}, a null QBER, needed for a successful forgery, is obtained with a probability upper-bounded, again, by the quantity $\big(\frac{3}{4}\big)^{\xi \lambda}$.

\end{enumerate}

\end{proof}

\begin{subtheorem}
\label{theorem_key_recycling}
    The secretly pre-shared integer-valued key $K_{\mathrm{Positions}}$ can be reused over an arbitrary number of QKD rounds.
\end{subtheorem}

\begin{proof}    
    On the one hand, measuring the sent bitstring by Alice alone does not give any information on $K_{\mathrm{Positions}}$. That is, such tampering with the transferred qubits can  only be used to gain information about $K_{\mathrm{Positions}}$ if combining it  with the knowledge of the  strings involved for encrypting  $B_{\mathrm{Alice}}$ and $B_{\mathrm{Bob}}$. 
    
    On the other hand, because $B_{\mathrm{Alice}}$ and $B_{\mathrm{Bob}}$ are selected uniformly at random, the encrypted public messages $\Tilde{B}_{\mathrm{Alice}}$ and $\Tilde{B}_{\mathrm{Bob}}$ reveal no information on the involved secret keys. Namely, as shown in \cite{ln_moser, shannon_equivocation}, the entropy $H(Z, k)$ associated to a secret binary key  $Z$ used $k$ times for XOR-ciphering a binary message M fulfills:

\begin{equation}
        H(Z, k) =  H(Z, 0) - k\gamma \mathrm{log}_2\mid \mathcal{Y} \mid,
\end{equation}
with $\mathcal{Y}$ being the space dimension of the secret key and where we have defined

\begin{equation}
        \gamma \equiv  1 - \frac{H(M)}{\mathrm{log}_2\mid \mathcal{Y} \mid},
\end{equation}     
     with $H(M)$ being the entropy associated to the message source.  Notice how $\gamma$ vanishes for uniformly random messages and that we are actually only interested in the case $k=1$. That is, the bases-encrypting keys are refreshed each round due to new quantum measurements, and we only need to make sure that no information is leaked in their single usage, because this could be combined\footnote{One could select a qubit position within the sent string and measure it in the computational basis, repeatedly across different rounds. This would potentially introduce no QBER, and if the outcomes obtained happened to be correlated with any bit learned from the bases-encrypting key, then it would follow that the selected qubit is an authenticating one.} with an intercept-resend strategy in order to learn $K_{\mathrm{Positions}}$. 
\end{proof}

In combination, proofs to Theorems \ref{theorem_theft}, \ref{theorem_classicical_theft} and \ref{theorem_key_recycling}, together with the proof to Proposition \ref{proposition_completeness}, serve as a proof for Theorem \ref{theorem_security}.
\end{proof}

\begin{proposition}
\label{proposition_noise}
    Under  the  \textit{acceptance condition a)} for  the  noise-adapted QKD authentication protocol defined in Section \ref{Noise_adaptation}, the probability $p_{\mathrm{A}}$ of a successful randomized forgery attempt fulfills:

    \begin{equation}
        p_{\mathrm{A}} \leq \frac{1}{2\xi\lambda}\frac{1}{\Big(\mu-\frac{1}{2}\Big)^2}.
    \end{equation}
\end{proposition}

\begin{proof}
    A randomized attacker, i.e., that one neither having  information about the outcomes of the entangled qubits involved nor having access to pre-shared and non-pre-shared secrecy, is  forced to acknowledge a bases choice, whatever part is the one suffering an impersonation attempt. In such case, the attacker can only guess with a probability of success of $\frac{1}{2}$ each of the bits obtained by the legitimate party. Each bit guess constitutes an event equivalent to guessing a fair coin toss, with $\frac{1}{2}$ being the expected value of its outcome and $\frac{1}{\sqrt{2}}$ its standard deviation. The central limit  theorem \cite{central_limit_theorem} together with Chebyshev's inequality \cite{chebyshev} ensures that the probability of obtaining a rate $1-QBER\geq\mu$,  with $\mu>\frac{1}{2} + \frac{1}{\sqrt{2\xi\lambda}}$, is

    \begin{equation}
       p_{\mathrm{A}} = p(1-QBER \geq \mu) \leq \frac{1}{2\xi\lambda}\frac{1}{\Big(\mu-\frac{1}{2}\Big)^2}.
    \end{equation}
\end{proof}

\subsection{AFC memory hardware parameters}
In Table.\,\ref{parameters:state}, we deliver an overview of the main hardware parameters and their evolution throughout the past recent years.

\subsection{Performance of the DNN}
\subsubsection{ROC Curve}
\begin{figure}
\includegraphics[width=0.45\textwidth,clip]{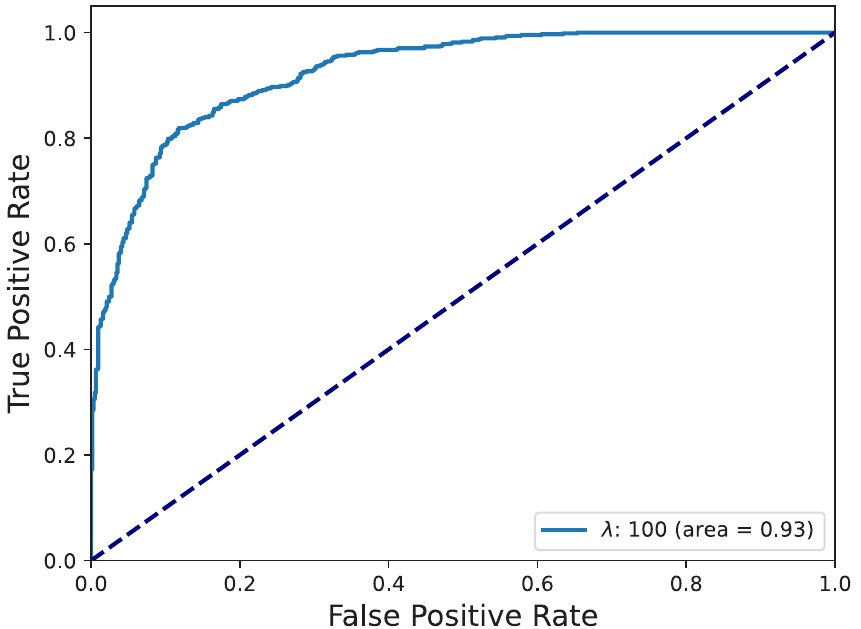}
\caption{ROC curve for $\lambda=100$.}

\label{Fig:prob3}       
\end{figure}

The Receiver Operating characteristic Curve (ROC) \cite{ROC} for the used DNN is shown in Fig.\,\ref{Fig:prob3}.

\subsubsection{Accuracy and cross entropy loss}
\begin{figure}
%\hspace*{0.4cm}
\includegraphics[width=0.5\textwidth,clip]{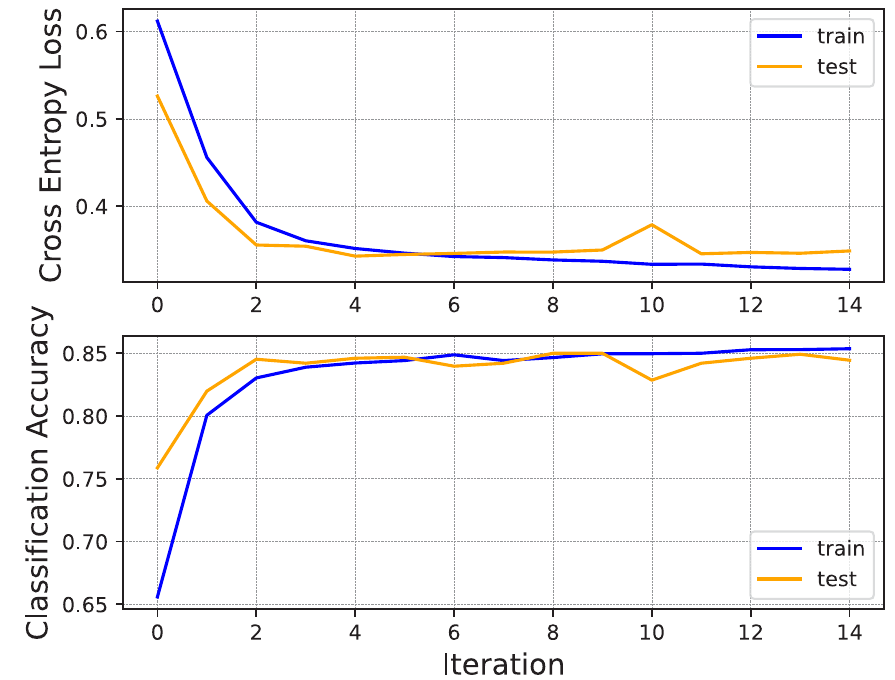}
\caption{Accuracy and cross entropy for the training and test set of $\lambda=100$ for the DNN, for distance of 1 $\mathrm{km}$ and storage time of $150$ $\mu \mathrm{s}$.}

\label{Fig:prob4}       
\end{figure}

The classification accuracy and cross-entropy loss are depicted in Fig.\,\ref{Fig:prob4}. The DNN achieves higher accuracy and lower cross-entropy loss during the training procedure. Fig.\,\ref{Fig:prob4} justifies the use of only 15 iterations for the DNN training, as we observe that both the cross-entropy loss and accuracy curves have stabilized by then. 

\end{document}